\pdfoutput=1
\makeatletter
\def\input@path{{./tex/}{./bst/}}
\makeatother
\documentclass[%
reprint,
superscriptaddress,
 amsmath,amssymb,
 aps,
 prl,
floatfix,
]{revtex4-2}

\usepackage{graphicx}
\usepackage{dcolumn}
\usepackage{bm}
\usepackage{hyperref}
\usepackage{siunitx}

\usepackage[dvipsnames]{xcolor}  

\newcommand{\mose}{MoSe$_2$}
\newcommand{\crbr}{CrBr$_3$}

\newcommand{\subfig}[2]{Fig.~\ref{#1}(#2)}
\newcommand{\panel}[1]{(#1)}

\newcommand{\abbrev}[3]{
  \newcounter{#3}
  \setcounter{#3}{0}
  \newcommand{#1}{\ifnum\value{#3}<1{#2 (#3)}\else{#3}\fi\stepcounter{#3}}}

\abbrev{\moke}{magneto-optical Kerr effect}{MOKE}
\abbrev{\PL}{photoluminescence}{PL}
\abbrev{\hBN}{hexagonal boron nitride}{hBN}
\abbrev{\mcd}{magnetic circular dichroism}{MCD}

\begin{document}


\title{Observation of Magnetic Proximity Effect Using Resonant Optical Spectroscopy of an Electrically Tunable MoSe$_2$/CrBr$_3$ Heterostructure}

\author{Livio Ciorciaro}
\email{livioc@phys.ethz.ch}
\affiliation{Institut f\"ur Quantenelektronik, ETH Z\"urich, Auguste-Piccard-Hof 1, 8093 Z\"urich, Switzerland.}
\author{Martin Kroner}%
\affiliation{Institut f\"ur Quantenelektronik, ETH Z\"urich, Auguste-Piccard-Hof 1, 8093 Z\"urich, Switzerland.}
\author{Kenji Watanabe}
\affiliation{National Institute for Materials Science, 1-1 Namiki, Tsukuba 305-0044, Japan}
\author{Takashi Taniguchi}
\affiliation{National Institute for Materials Science, 1-1 Namiki, Tsukuba 305-0044, Japan}
\author{Atac Imamoglu}
\affiliation{Institut f\"ur Quantenelektronik, ETH Z\"urich, Auguste-Piccard-Hof 1, 8093 Z\"urich, Switzerland.}

\date{\today}

\begin{abstract} Van der Waals heterostructures combining two-dimensional magnetic and
semiconducting layers constitute a promising platform for interfacing magnetism, electronics, and
optics. Here, we use resonant optical reflection spectroscopy to observe the magnetic proximity effect
in a gate-tunable MoSe$_2$/CrBr$_3$ heterostructure. High quality of the interface leads to a giant
zero-field splitting of the K and K' valley excitons in MoSe$_2$, equivalent to an external magnetic
field of \SI{12}{T}, with a weak but distinct electric field dependence that hints at potential for
electrical control of magnetization. The magnetic proximity effect allows us to use resonant optical
spectroscopy to fully characterize the CrBr$_3$ magnet, determining the easy-axis coercive field,
the magnetic anisotropy energy, and critical exponents associated with spin susceptibility and
magnetization. 
\end{abstract}

\maketitle


Two-dimensional (2D) magnetic materials have attracted considerable attention due to their potential
applications in spintronic devices \cite{Gong2019,Cortie2019}. Since the first demonstration that
magnetism persists down to the monolayer limit in chromium trihalides (CrX$_3$, X = Cl, Br,
I)\cite{Huang2017,Zhang2019a}, much progress has been made, both in understanding fundamental
properties of these materials
\cite{Zhong2017,Chen2018,Jin2018,Thiel2019,Jin2019,Song2019,Cai2019,Chen2019a,Sun2019,Zhong2020} and
investigation of crucial steps towards applications
\cite{Wang2018b,Song2018,Ghazaryan2018,Klein2018,Jiang2018b,Seyler2018,Huang2018,Jiang2018a,Farooq2019a}.
Concurrently, transition metal dichalcogenides (TMDs) have established themselves as 2D
semiconductors with remarkable optical properties \cite{Xu2014,Mak2018,Onga2017} and possible
applications in photonics and valleytronics \cite{Mak2010,Manzeli2017,Deng2018}. Van der Waals
heterostructures composed of different 2D materials have the potential to realize atomically smooth
interfaces that are not affected by lattice structure mismatch between the layers, allowing in
principle arbitrary combinations of materials \cite{Geim2013}. Magnetic proximity effect in such
structures on the one hand leads to transfer of magnetization to otherwise non-magnetic layers, and
on the other hand may allow for controlling magnetization using electrical or optical excitation.

In this Letter, we use resonant optical spectroscopy to unequivocally demonstrate the magnetic proximity
effect in a \mose/\crbr{} heterostructure, where we observe a large zero-field splitting of the K
and K' exciton resonances in \mose{}. We find that the magnetization of \mose{} is exclusively
induced by exchange coupling of conduction band electrons. We use the shift of \mose{} excitonic
resonances to study the magnetic properties of \crbr{}, and determine the magnetic
anisotropy as well as the critical exponents associated with magnetization and susceptibility. Our
work establishes resonant optical measurements in heterostructures incorporating TMD monolayers and
2D magnetic materials as a powerful spectroscopic tool that could be invaluable for studying
magnetic materials with weak optical transitions without requiring high power laser excitation.

\begin{figure}
    \includegraphics{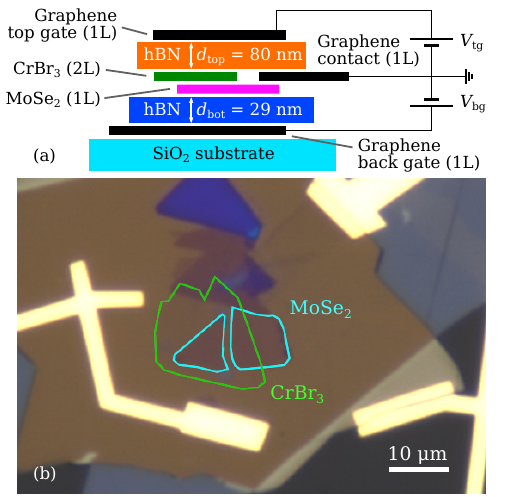}
    \caption{(a) Schematic of the layer structure of the device and electrical connectivity.
             A bilayer \crbr{} and monolayer \mose{} are encapsulated in hBN. Monolayer graphene flakes
             are used as top and bottom gates and as contact to \mose. The stack is placed on a
             transparent SiO$_2$ substrate. (b) Optical micrograph of the device, with \mose{}
             and \crbr{} outlined in blue and green, respectively.}
    \label{fig:sample}
\end{figure}

The sample we studied consists of a monolayer \mose{} in direct contact with a bilayer \crbr{}
encapsulated in \hBN{} on a SiO$_2$ substrate, as shown schematically in \subfig{fig:sample}{a}.
In contrast to the layer-antiferromagnets CrI$_3$ and CrCl$_3$, the interlayer exchange in
bilayer CrBr$_3$ has been shown to be ferromagnetic \cite{Zhang2019a}.
Monolayer graphene gates and a graphene contact allow for independent tuning of the charge carrier
density and out-of-plane electric field in the sample. In the optical micrograph in
\subfig{fig:sample}{b}, the regions of bare \mose, bare \crbr, and the overlapping region can be
seen. Details on the sample fabrication, optical setup, and data analysis are given in the
Supplemental Material \cite{supplemental}. All measurements were performed at approximately
\SI{6}{K} unless stated otherwise.

\begin{figure}
    \includegraphics{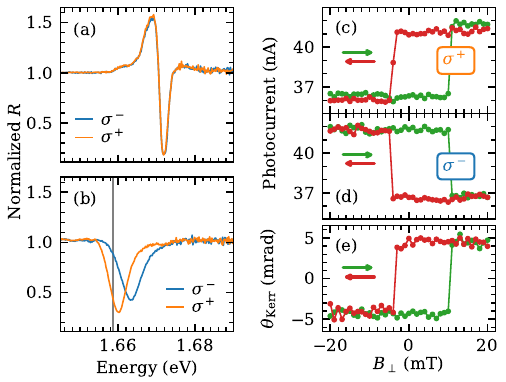}
    \caption{(a) Reflection spectrum of bare undoped \mose{}. (b) Reflection spectra in
             $\sigma^-$ and $\sigma^+$ polarization of the undoped \mose/\crbr{} heterostructure at
             $B = 0$. The K and K' valley excitons are split by \SI{2.9}{meV}. (c), (d) Hysteresis
             of the reflection of a $\sigma^+$/$\sigma^-$ polarized laser tuned to the low-frequency
             tail of the \mose{} exciton in the heterostructure as marked by the vertical line in
             panel (b). (e) Measurement of the \crbr{} magnetization hysteresis using the MOKE.
             The apparent offset from zero of the magnetic field in both measurements is due
             to a stray field from a ferromagnetic component in the cryostat.}
    \label{fig:splitting}
\end{figure}

Normalized polarization-resolved reflection spectra of the bare \mose{} and the \mose/\crbr{}
heterostructure are shown in \subfig{fig:splitting}{a} and \panel{b}, for a choice of gate voltages
that ensure charge neutrality of \mose{}. In the absence of an external magnetic field, the K and K'
valley excitons are degenerate in bare \mose{} and the spectrum shows no polarization dependence. In
contrast, a valley splitting of \SI{2.9}{meV} emerges in the \mose/\crbr{} heterostructure region,
equivalent to an external magnetic field of \SI{12}{\tesla}, assuming the electronic g-factor to be
$4$. This splitting can be attributed to an exchange coupling between electronic states in \mose{}
and spin-polarized states in \crbr{} that leads to different energy shifts for the \mose{} K and K'
valley excitons. Due to strain and disorder in the heterostructure, the splitting varies
spatially by approximately \SI{10}{\%}.

To demonstrate that the \mose{} exciton valley splitting originates from the magnetization of
\crbr{}, we compare hysteresis measurements of the \mose{} reflectance as a function of an external
out-of-plane magnetic field $B_\perp$ to the established method of measuring \crbr{} magnetization along the
easy axis using the \moke{} \cite{Zhang2019a}. Figures~\ref{fig:splitting}\panel{c} and \panel{d}
show the reflectance of a right- and left-hand circularly polarized ($\sigma^+$/$\sigma^-$) laser
(\SI{5}{\micro\watt} CW) tuned to the low-frequency
tail of the exciton resonance, as indicated by the vertical line in \subfig{fig:splitting}{b}, as a
function of $B_\perp$. For this choice of laser detuning, the valley splitting gives rise to maximal
contrast in \mcd. The \moke{} depicted in \subfig{fig:splitting}{e} is measured on the same spot
with a linearly polarized laser at \SI{2.755}{eV} (\SI{450}{nm}, \SI{20}{\micro\watt} CW).
The one-to-one correspondence between the
measurements confirms that the valley splitting is directly linked to the magnetization of \crbr.
We verified that the \moke{} signal is not altered by the presence of \mose{} by comparing
measurements on the heterostructure and bare \crbr{} (see Supplemental Material \cite{supplemental}).
Using resonant spectroscopy on \mose{} instead of the \moke{} to access the magnetization of \crbr{}
is advantageous, since it allows us to perform the same measurement with a simpler technique and
lower illumination power. Avoiding measurements requiring high laser intensity is particularly
important for chromium trihalides where sizeable \moke{} signals are only obtained using
above-band-gap lasers that could cause heating. Moreover, identifying peak positions instead of
measuring laser intensities after a polarizing beam splitter makes our spectroscopic method less
sensitive to imperfections in the polarization selection than traditional techniques.

An exchange splitting of similar magnitude has been reported in the pioneering work on
\PL{} measurements of WSe$_2$/CrI$_3$
heterostructures \cite{Zhong2017,Seyler2018,Zhong2020}. We were not able
to observe the splitting in \PL{} measurements (see Supplemental Material \cite{supplemental}).
Instead, we see broad emission lines
with an integrated intensity that is smaller by a factor twenty compared to bare \mose{},
which suggests that tunneling to \crbr{} provides a fast non-radiative relaxation channel for
conduction band electrons and excitons in \mose{}. Because the exchange coupling responsible for
the splitting relies on second-order virtual tunnel coupling, the PL splitting is expected to be large where 
the tunnel coupling is large, leading to a short exciton lifetime. Since PL primarily
originates from long-lived states, we would expect it to be dominated by low-oscillator-strength
localized excitations in parts of the heterostructure where the tunnel coupling is small.
Consequently, disorder-induced spatial variations of the tunnel coupling could lead to a \PL{}
signal that shows small or possibly vanishing exciton valley splitting. This is in contrast to
resonant reflection/absorption measurements which probe extended states with high oscillator
strength within the optical excitation spot.

\begin{figure}
    \includegraphics{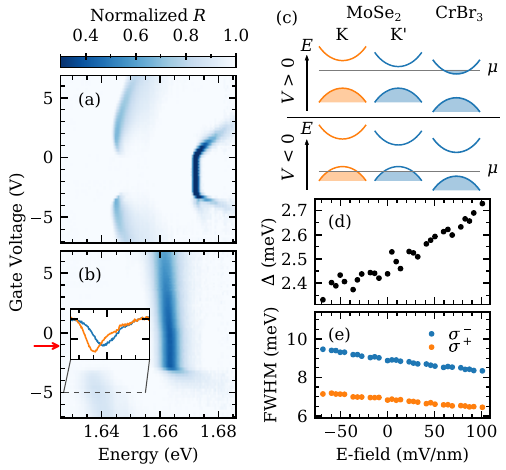}
    \caption{(a) Gate dependence of the bare \mose{} reflection spectrum. The appearance of 
             attractive and repulsive polaron lines at positive and negative voltages shows the
             flake can be n- and p-doped. (b) Gate dependence of the reflection spectrum of the
             heterostructure in $\sigma^-$ polarization. Due to the type-II band alignment, \mose{}
             cannot be n-doped anymore, as signified by the persistence of the exciton line at
             positive voltages.
             Inset: Reflection spectra in the p-doped regime, as indicated by the dashed line.
             The attractive polaron lines show the same splitting as the exciton lines.
             The x-ticks are identical to the parent axis, the y-tick separation is 0.1.
             (c) Schematic of the type-II band aligment of \mose{} and \crbr{} with
             chemical potential $\mu$ for positive and negative gate voltages.
             Exchange coupling leads to valley splitting in the conduction band of \mose{}.
             (d), (e) Dependence of the valley splitting $\Delta$ and FWHM of the exciton lines on the
             out-of-plane electric field. Both the splitting and FWHM are tunable, indicating a modification
             of the tunneling rate across the \mose/\crbr{} interface.
    }
    \label{fig:gate_dep}
\end{figure}

To explore the nature of the exchange coupling between \crbr{} and \mose{} we measure the gate
voltage dependence of the reflection spectra in $\sigma^-$ polarization, shown in
\subfig{fig:gate_dep}{a} and \panel{b}. The voltages indicated on the vertical axis were applied to
both top and bottom gate. The bare \mose{} flake can be charged with electrons or holes, evidenced
by the appearance of attractive polaron lines at both positive and negative gate voltages
\cite{Sidler2016}. In the presence of \crbr{}, only holes can be injected into \mose{}, consistent
with the type-II band alignment schematically shown in \subfig{fig:gate_dep}{c} and
predicted from ab initio calculations \cite{Zollner2019,Xie2018}. Injected
electrons accumulate in the lower-lying conduction band of \crbr, leaving the \mose{} undoped and
leading to screening of the top gate (see Supplemental Material \cite{supplemental}).

The attractive polaron line on the p-doped side exhibits the same valley splitting as the neutral
exciton, as shown in the inset of \subfig{fig:gate_dep}{b}. If the itinerant holes in \mose{} were
subject to a sizeable exchange interaction, we would observe a strong valley polarization of holes,
leading to a single circularly polarized attractive polaron resonance \cite{Back2017,Smolenski2019}.
The observation of polaron resonances with equal strength for the two polarizations demonstrates that
electron exchange is predominantly responsible for the exciton and polaron valley splitting.
A detailed understanding of the underlying coupling mechanism 
is beyond the scope of this Letter and requires additional theoretical work.

The presence of top and bottom gates allows us to probe the electric field dependence of the
reflectance for constant chemical potential. In the absence of mobile charges in the
heterostructure, the electric field is approximately given by
$E \approx \left(V_\mathrm{tg} - V_\mathrm{bg}\right) / \left(d_\mathrm{top} + d_\mathrm{bot}\right)$
and the chemical potential by
$\mu \propto d_\mathrm{bot} V_\mathrm{tg} + d_\mathrm{top} V_\mathrm{bg}$,
where $d_\mathrm{top}$ and $d_\mathrm{bot}$ are the the thicknesses of the top and bottom \hBN{}
flakes; the actual electric field may deviate by a constant factor due to the dielectric
constants and finite thickness of \mose{} and \crbr{} (see Supplemental Material \cite{supplemental}).
To keep the chemical potential constant
while varying the electric field, we tune the gate voltages with a fixed ratio
$(V_\mathrm{tg} - V_0)/V_\mathrm{bg} = 3.3$, where $V_0$ determines the chemical potential;
we determined this ratio experimentally from 2D gate sweeps (see Supplemental Material \cite{supplemental}).

Figures~\ref{fig:gate_dep}\panel{d} and \panel{e} show the valley splitting and the reflection peak
widths, respectively, for the neutral exciton as a function of the applied electric field. The
choice of gate voltages, indicated by the red arrow in \subfig{fig:gate_dep}{b}, ensures the charge
neutrality of the heterostructure. Clear dependence of both the splitting and the linewidth on the electric field suggests
that the tunnel coupling strength is modified. Such an approximately linear electric field
dependence of the splitting was predicted in theoretical works \cite{Zollner2019,Xie2018} and may
have implications for future gate-tunable spintronic devices. Additionally, the higher-energy
exciton line (here $\sigma^-$) is consistently broader than the lower-energy line, presumably due to
the spin-dependent charge transfer between \mose{} and \crbr{}; similar observations were previously
reported in heterostructures composed of different 2D magnetic layers \cite{Seyler2018,Zhong2020}.

\begin{figure}
    \includegraphics{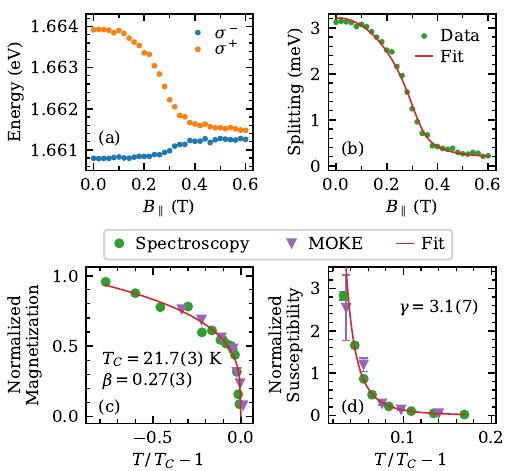}
    \caption{(a) Fitted exciton peak positions from polarization-resolved reflection spectra as
             function of in-plane magnetic field. The splitting collapses around \SI{0.3}{\tesla},
             indicating the transition to in-plane magnetization of \crbr.
             (b) Exciton valley splitting as function of in-plane magnetic field. The fitted model
             using a macroscopic magnetic moment with anisotropy is in excellent agreement with the
             experimental data.
             (c) Magnetization of \crbr{} as function of temperature. Round data points are
             extracted from polarization-resolved reflection spectra of \mose{} and triangular data
             points are extracted from MOKE measurements.  A fit of the form
             $m(T) = A \left(1 - T/T_\mathrm{C}\right)^\beta$ yields values for the critical
             temperature $T_\mathrm{C}$
             and critical exponent $\beta$. 
             (d) Magnetic susceptibility as function of temperature extracted from spectra and MOKE
             measurements. A fit of the form
             $\chi(T) = B \left(T/T_\mathrm{C} - 1 \right)^{-\gamma}$ yields the critical exponent
             $\gamma$.
             }
    \label{fig:criticality}
\end{figure}

Having demonstrated the magnetic proximity effect in \mose{}, we use resonant spectroscopy of the
\mose{} exciton resonance to determine the magnetic properties of \crbr.
Figure~\ref{fig:criticality}\panel{a} shows fitted positions of the split exciton peaks as function of an
applied in-plane magnetic field $B_\parallel$. We observe that for $B_\parallel \ge \SI{0.1}{T}$,
the splitting gradually decreases and saturates at a value of \SI{0.1}{meV} for \SI{0.3}{\tesla}.
The reduction of the splitting is a consequence of the canting of the \crbr{} spins into the plane.
The small remaining splitting at high magnetic fields is due to a tilt of the magnetic field axis
with respect to the sample plane that leads to an out-of-plane component of the magnetization and
consequently a non-zero exchange field. A striking feature of the data in
\subfig{fig:criticality}{a} is the asymmetry in the $B_\parallel$-induced change in resonance energy
between the low- and high-energy exciton peaks: we speculate that this asymmetry could arise from an
energy splitting between the spin-polarized conduction bands of \crbr{} that play a prominent role
in determining the exchange coupling to \mose{} electrons in different valleys.

The observed dependence of the splitting vs.\ $B_\parallel$ allows us to estimate the
anisotropy energy of \crbr. To this end, we assume that the \crbr{} flake has a uniform magnetization $\vec{m}$ and numerically minimize its potential energy
$E = K \left(\sin\theta\right)^2 + \left|m\right| B_\parallel \cos(\theta - \alpha)$,
where $K$ is the anisotropy energy along the easy axis, $\theta$ is the angle of the magnetic moment
with respect to the easy axis, $B_\parallel$ is the external magnetic field, and $\alpha$ is the
angle of $B_\parallel$ with respect to the easy axis. We set $\left|m\right| =
3.87\mu_\mathrm{B}$ to the magnetization per Cr atom \cite{Kotani1949,Dillon1962} and fit the model
to the experimental data to obtain an anisotropy energy $K = \SI{34.3(4)}{\micro eV}$ per Cr atom
and $\alpha = \SI{88.20(13)}{\degree}$. Although this simple model does not tell us anything about
the type of anisotropy, previous calculations have shown that it is expected to originate from
anisotropic exchange coupling rather than on-site anisotropy \cite{Lado2017}. By using the
previously reported values of the isotropic intralayer exchange of bulk \crbr{} ($J \approx
\SI{0.8}{meV}$) \cite{Davis1964,Samuelsen1971}, we find that the exchange interaction is weakly
anisotropic, $J / K \approx 20$. The small coercive field we measure
(\subfig{fig:splitting}{c}--\panel{e}) is consistent with the weak anisotropy of the intralayer exchange
interaction. Remarkably, the exciton valley splitting in \mose{} is comparable to the \crbr{}
intralayer exchange $J$, even though the exchange between two \crbr{} layers is expected to be
significantly smaller \cite{Davis1964,Samuelsen1971,deJongh1974}.

Next, we measure the critical temperature $T_C$ and critical exponents of the second-order magnetic phase
transition of \crbr{} through resonant \mose{} exciton spectroscopy. To this end, we measure the hysteresis
curves of circularly polarized reflection spectra vs.\ $B_\perp$ as a
function of temperature (see Supplemental Material \cite{supplemental}).
For $T < T_\mathrm{C}$, shown in \subfig{fig:criticality}{c}, the splitting at $B_\perp = 0$
provides a measure for the remnant magnetization $m$, which is the order parameter for the phase transition. 
For $T > T_\mathrm{C}$, shown in \subfig{fig:criticality}{d}, the slope of the splitting vs.\ 
$B_\perp$ is proportional to the magnetic susceptibility $\chi$.
By fitting the functional forms $m(T) = A_1 \left(1 - T/T_\mathrm{C}\right)^\beta$ and
$\chi(T) = A_2 \left(T/T_\mathrm{C} - 1\right)^{-\gamma}$ simultaneously to the experimental data,
we find the Curie temperature $T_\mathrm{C} = \SI{21.7(3)}{K}$ as well as the critical exponents
$\beta = 0.27(3)$ and $\gamma = 3.1(7)$. We also perform
similar measurements using the \moke{} and find that the data points fall onto the same curves when
normalized with respect to the peak values of $m$ and $\chi$. The values we obtain for $\beta$ and
$\gamma$ are consistent with the 2D-Heisenberg model with weak anisotropy \cite{Fisher1974}. 

In conclusion, we use resonant exciton reflection measurements of valley splitting in \mose{} to
demonstrate a strong magnetic proximity effect due to ferromagnetic \crbr. From the
absence of itinerant hole valley polarization in the
reflection spectra we infer that the resulting valley Zeeman effect is predominantly due to
exchange coupling between conduction band electrons in \mose{} and \crbr: remarkably, the strength of
this interlayer exchange coupling is comparable to the intralayer
exchange coupling in \crbr.
Our investigation of the magnetic
properties of \crbr{} using resonant optical spectroscopy reveals several features such as
an electric field dependence of the proximity effect, weak anisotropy
$J / K \approx 20$ of the exchange interaction, and the critical exponents associated with the magnetic phase transition.

The data that supports the findings of this Letter is available in the ETH Research Collection \cite{ethResearchCollection}.

\section{Acknowledgments}
We gratefully acknowledge the contributions of Alberto Morpurgo and Zhe Wang who introduced us to
the field of 2D magnets. Y.\ Shimazaki and A.\ Popert contributed to the fabrication of the
heterostructure. We also express our gratitutde to A.\ Vindigni and P.\ Maletinsky for insightful
discussions. This work was supported by the Swiss National Science Foundation (SNSF) under Grant No. 200021-178909/1. K.~W.\ and T.~T.\ acknowledge support from the Elemental Strategy
Initiative conducted by the MEXT, Japan, and the CREST (JPMJCR15F3), JST.

\nocite{*}

\bibliographystyle{apsrev4-2}
\bibliography{references,suppl}

\end{document}



\title{Supplemental Material: Observation of Magnetic Proximity Effect in Resonant Optical Spectroscopy of an Electrically Tunable MoSe$_2$/CrBr$_3$
       Heterostructure}

\author{Livio Ciorciaro}
\email{livioc@phys.ethz.ch}
\affiliation{Institut f\"ur Quantenelektronik, ETH Z\"urich, Auguste-Piccard-Hof 1, 8093 Z\"urich, Switzerland.}
\author{Martin Kroner}%
\affiliation{Institut f\"ur Quantenelektronik, ETH Z\"urich, Auguste-Piccard-Hof 1, 8093 Z\"urich, Switzerland.}
\author{Kenji Watanabe}
\affiliation{National Institute for Materials Science, 1-1 Namiki, Tsukuba 305-0044, Japan}
\author{Takashi Taniguchi}
\affiliation{National Institute for Materials Science, 1-1 Namiki, Tsukuba 305-0044, Japan}
\author{Atac Imamoglu}
\affiliation{Institut f\"ur Quantenelektronik, ETH Z\"urich, Auguste-Piccard-Hof 1, 8093 Z\"urich, Switzerland.}

\date{\today}

\maketitle

\section{Sample Fabrication}
All flakes were exfoliated from crystals onto Si/\sio\ substrates, and thicknesses identified by their
optical contrast. The \hBN{} crystals were provided by our collaborators at NIMS and all other crystals
were purchased from HQ Graphene. The stack was
assembled using a standard dry transfer technique \cite{Wang2013} with poly(bisphenol A carbonate)
and deposited onto a 0.5-mm-thick single-crystal quartz substrate.
Electrodes were fabricated using standard lithography techniques.

\section{Optical Setup}
\begin{figure}
\includegraphics{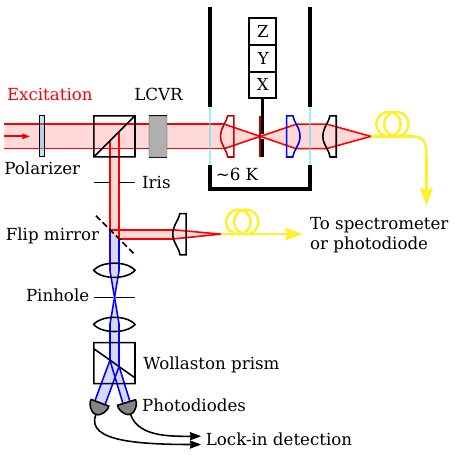}
\caption{Schematic of the optical setup. The circular polarization was set using a liquid crystal
         variable retarder (LCVR).}
\label{fig:setup}
\end{figure}

A schematic of the optical setup is shown in \fig{fig:setup}.
The measurements were carried out in a confocal microscope. For reflection measurements, an Exalos
superluminescent LED centered at \SI{1.631}{eV} (\SI{760}{nm}) was used, and a \SI{1.797}{eV}
(\SI{690}{nm}) laser diode and a HeNe-laser
(\SI{1.959}{eV}, \SI{632.8}{nm}) were used as pump in \PL{} measurements (both pump lasers yielded the same signal).
The measurements of the induced circular dichroism in \mose{} (Fig.~2(c) and (d) of the main text) were
done with a tunable continuous wave Ti:sapphire laser detected with a photodiode.
The \moke{} measurements were done with a \SI{2.755}{eV} (\SI{450}{nm}) laser diode
and the intensity was modulated using an acousto-optic modulator for lock-in detection.
Single aspheric lenses with \NA{} of 0.7 (red light)
and 0.55 (blue light) were used to focus the lasers on the sample. The two lenses were mounted
simultaneously inside the cryostat on the front side (red) and back side (blue) of the sample, such that
Kerr rotation and spectra could be measured by rotating the sample by 180 degrees without warming up.
The Kerr rotation was measured through the substrate. To avoid effects from the birefringence of the
quartz substrate, the linear polarization was chosen along a crystal axis.

\section{Temperature Tuning}
In the cryostat where the measurements were done, the sample is cooled directly by a flow of helium gas in
a variable temperature insert.
The temperature can be tuned by heating the helium gas at the inlet with a resistive heater.
To measure the sample temperature we mounted an Allen-Bradley resistor on the sample holder
at approximately the same height as the sample along the temperature gradient. Its resistance curve as
a function of temperature was measured in advance using a different system with a calibrated thermometer
and fitted using the three-parameter model \cite{Kopp1972}
\begin{equation}
\log R = A + \frac{B}{T^P}.
\end{equation}
Differences in the resistance of the lines and a temperature gradient between the sample and the resistor
may lead to systematic errors in the temperature readout. However, in the relevant range of temperatures
around the critical temperature these would lead to a nearly linear shift of the temperature axis and would
not affect the critical exponents significantly.

\section{Gate Tuning and hBN Thickness}
\label{sec:hBN_thickness}
\begin{figure}
\includegraphics{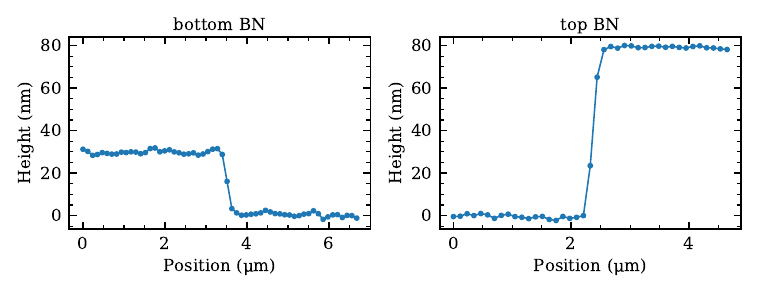}
\caption{Height profiles of the bottom and top hBN flakes measured with AFM.}
\label{fig:AFM}
\end{figure}

The thicknesses of the \hBN{} flakes were estimated from their colors and verified with \afm{}
measurements, shown in \fig{fig:AFM}.
As discussed in the main text, the thickness ratio of the two \hBN{} flakes determines the voltage
ratio needed between top and bottom gate to apply an electric field $E$ without changing the
chemical potential $\mu$. In a first approximation we consider the parallel plate capacitor formed by the top
and bottom gates to be filled uniformly
with a dielectric medium with relative permittivity $\varepsilon_\mathrm{rel}$, thus neglecting the 
finite thicknesses of the \crbr{} bilayer and \mose{} monolayer. Then we have
$E = \left(V_\mathrm{tg} - V_\mathrm{bg}\right) / \left(d_\mathrm{top} + d_\mathrm{bot}\right)$
and
$\mu \propto d_\mathrm{bot} V_\mathrm{tg} + d_\mathrm{top} V_\mathrm{bg}$, as defined in the main text.
This means that in order to tune the electric field while keeping the chemical potential constant,
the applied gate voltages need to satisfy the relation
$V_\mathrm{tg} = \alpha V_\mathrm{bg} + V_0$, where $\alpha = d_\mathrm{top} / d_\mathrm{bot} = 2.75$
and the offset $V_0$ determines the chemical potential. In practice we find that the correct
slope is $\alpha = 3.3$, by measuring a 2D
sweep of top and bottom gate voltages (see Supplemental Sec.~\ref{sec:gate_dep}) and looking at the slope of the line
where the sample becomes $p$-doped and the exciton disappears. This discrepancy is due to the finite thicknesses
and different relative permittivities of \mose{} and \crbr{} compared to \hBN.
Taking these into account the electric field is given by
\begin{displaymath}
E = 
\frac{V_\mathrm{tg} - V_\mathrm{bg}}{d_\mathrm{bot} + d_\mathrm{top}} \cdot
\left(\frac{\varepsilon_\mathrm{MoSe}}{\varepsilon_{h\text{-BN}}}
+ \frac{\varepsilon_\mathrm{MoSe}}{\varepsilon_\mathrm{CrBr}} \frac{d_\mathrm{CrBr}}{d_\mathrm{bot} + d_\mathrm{top}}
+ \frac{d_\mathrm{MoSe}}{d_\mathrm{bot} + d_\mathrm{top}} \right)^{-1}.
\end{displaymath}
While the values will be different depending on the ratio of dielectric constants, the field is still
proportional to the difference between the applied top and bottom gate
voltages as long as there are no free charges in the heterostructure (between the capacitor plates),
which we can verify spectroscopically.

\section{Bare CrBr$_3$ MOKE}
\begin{figure}
\includegraphics{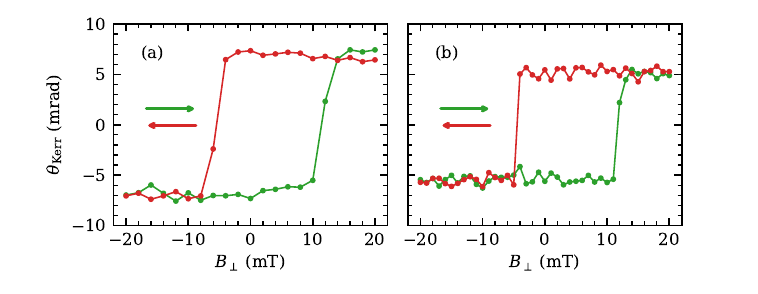}
\caption{MOKE measured on (a) bare \crbr{}, and (b) the heterostructure.}
\label{fig:moke_comparison}
\end{figure}

To verify that we indeed probe the \crbr{} magnetization when measuring the \moke{} of the
heterostructure, we compare them to \moke{} measurements of bare \crbr.
Figures~\ref{fig:moke_comparison}(a) and (b) show the \moke{} measured on bare \crbr{} and on the
heterostructure, respectively. The signal is very similar, with small differences in magnitude of
the rotation angle and coercive field which can be explained by sensitivity on the focus position and
disorder in the sample.

\section{Photoluminescence Measurements}
\begin{figure}
\includegraphics{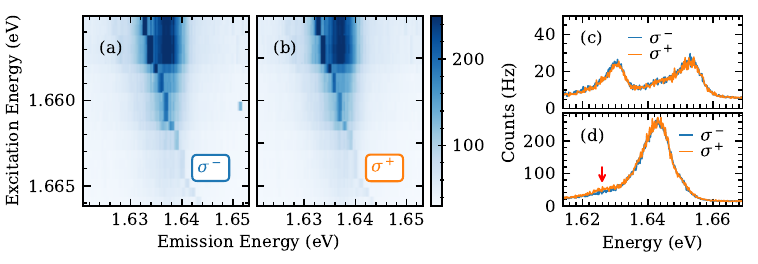}
\caption{(a), (b) Photoluminescence spectrum vs.\ exciation wavelength in $\sigma^-$ and $\sigma^+$
         detection under linearly polarized excitation. There is no difference between the
         two polarizations in the detected light. The sharp peak that depends linearly on the
         excitation wavelength is a Raman peak from the optical fiber.
         (c) A typical PL spectrum of the heterostructure. There is no measureable difference between the
         polarizations.
         (d) The PL spectrum at a particular spot near the edge of the \mose{} flake shows a
         slight difference in emission intensity between the two circular polarizations, but no splitting.}
\label{fig:PL}
\end{figure}

As stated in the main text, we do not observe a consistent polarization splitting of the \PL{} signal of the
heterostructure. Even in photoluminescence excitation measurements under linear excitation of the exciton
peak and circular detection we do not observe a splitting or any clear polarization dependence, as
shown in \subfig{fig:PL}{a} and \subfigalt{b}. The sharp line that moves linearly with the
excitation wavelength is a Raman peak stemming from the optical fiber. A typical PL spectrum without
any polraization dependence is shown in \subfig{fig:PL}{c}. Near the edges of the sample
some spots can be found where there is a small intensity difference between the two circular
polarizations on the red tail of the broad emission peak. An example is shown in
\subfig{fig:PL}{d}. This polarization dependence also responds to a switching of the \crbr{}
magentization. However, since we observe this behavior only on isolated spots near the edge of the
flake, we do not consider it a generic feature that is relevant to our discussion.

\section{Fast Charge Transfer: Optical Doping}
\begin{figure}
\includegraphics{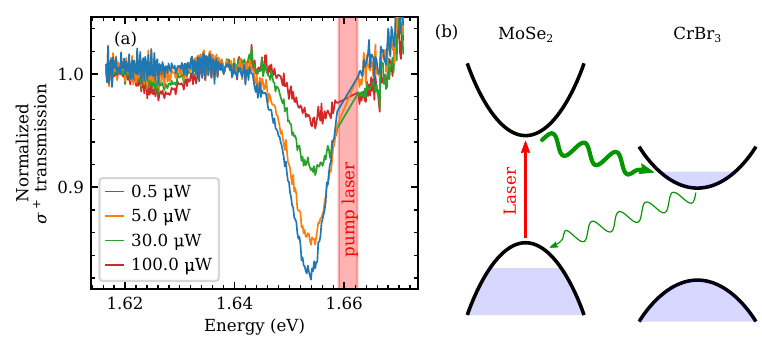}
\caption{(a) Transmission spectra of the heterostructure measured while pumping the exciton resonance
         with different powers. As the pump power increases, the exciton peak broadens and blueshifts,
         while the attractive polaron peak emerges, signifying that we optically dope \mose.
         (b) Schematic of relaxation dynamics in the heterostructure under laser excitation of \mose.
         Fast decay of electrons from the \mose{} conduction band into long lived states in the \crbr{}
         conduction band lead to a dynamical equilibrium with an excess hole population in \mose.}
\label{fig:opt_doping}
\end{figure}

In the main text we argue that the broadening of the exciton lines of the heterostructure with respect
to bare \mose{} excitons as well as the weak \PL{} signal stem from fast tunneling of electrons
from the \mose{} conduction band to the \crbr{} conduction band. To support this claim and to exclude that
the linewidth is simply dominated by inhomogeneous broadening we show that we can optically dope
\mose{} in the presence of \crbr. For this, we measure the transmission spectrum in circular polarization
at zero gate voltage while pumping the high-frequency tail of the exciton resonance with a laser in the
orthogonal polarization. Transmission spectra at different pump powers are shown in
\subfig{fig:opt_doping}{a}. At pump powers above \SI{30}{\micro\watt}, the exciton resonance
blueshifts and broadens considerably and the attractive polaron resonance emerges.
This is consistent with fast non-radiative electron relaxation, leaving behind itinerant holes
when the photoexcited electrons tunnel out of \mose, as shown schematically in
\subfig{fig:opt_doping}{b}. At the same pump powers on bare \mose{} we do not see the attractive polaron
peak emerge. Instead, \PL{} from the \mose{} exciton starts to dominate the spectrum as the pump power
increases.

\section{Extended Gate Dependence Data}
\label{sec:gate_dep}
\begin{figure}
\includegraphics{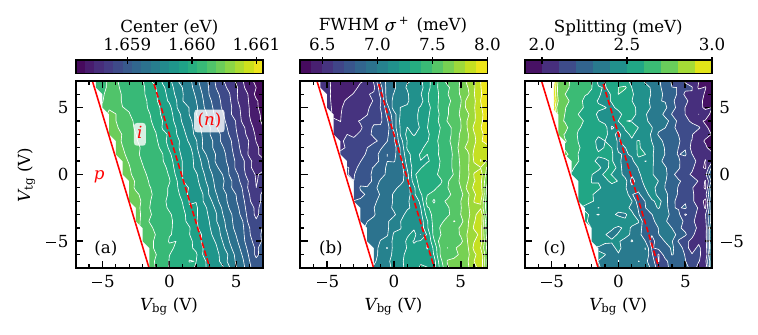}
\caption{(a) Contour plot of the average of the fitted exciton peak positions from circularly
         polarized reflection spectra of the heterostructure as function of both gate voltages. In
         the voltage ranges marked $p$ and $i$, \mose{} is $p$-doped or neutral, respectively, and
         in the region marked $(n)$ the bare \mose{} is $n$-doped, providing electrical contact to
         the \crbr{} flake. The exciton line redshifts as the \crbr{} flake becomes charged.
         (b) Contour plot of the fitted FWHM of the $\sigma^+$ exciton as function of gate
         voltages. The peak width is sensitive only to the vertical electric field.
         Vertical contour lines in the $(n)$ region indicate that the charged \crbr{} flake
         screens the \mose{} flake from the top gate.
         (c) Contour plot of the fitted valley splitting as function of gate voltages. Similar
         to the FWHM, the splitting is affected mostly by the vertical electric field.}
\label{fig:gate_dep}
\end{figure}

In \fig{fig:gate_dep} we show the full data of reflection spectra vs.\ top and bottom gate voltage.
Contour plots of fitted parameters of the exciton resonances as function of both gate voltages are
shown in \subfig{fig:gate_dep}{a}--\subfigalt{c}. Voltage ranges labeled $p$ and $i$ correspond to
$p$-doped and neutral \mose{} throughout the whole heterostructure, while in the region labeled $(n)$
only the bare \mose{} is $n$-doped, creating an electrical contact to the \crbr{} flake and allowing
it to be $n$-doped as well. The straight red lines separating the regions also represent lines of constant
chemical potential (see Supplemental Sec.~\ref{sec:hBN_thickness}).
Figure~\ref{fig:gate_dep}\subfigalt{a} shows the gate dependence of the center wavelength of
the two exciton peaks (average of the two peak positions). The direction of the contour lines
indicates that the redshift depends only
on the chemical potential and we attribute this shift to a weak attractive interaction between an
exciton in \mose{} and the electrons populating the \crbr{} conduction band when it is charged. In
\subfig{fig:gate_dep}{b} we plot the FWHM of the $\sigma^+$-polarized resonance.  In the neutral
region the contour lines indicate that the peak width depends mostly on the electric field. In the
$(n)$ region \crbr{} becomes charged and screens the \mose{} flake from the top gate, as evidenced by
the vertical orientation of the contour lines, making it impossible to independently control electric
field and chemical potential. The $\sigma^-$-polarized resonance shows the same qualitative behavior. The
splitting between the excitons also shows a similar trend as the widths, as shown in
\subfig{fig:gate_dep}{c}.

\section{Data Analysis}
\subsection{Spectroscopy}

The reflection and transmission spectra were normalized by taking spectra next to the \mose{}
and \crbr{} flakes, but still on the stack of graphene--\hBN--\hBN--graphene. To reduce the
sensitivity on focal drifts when moving the nanopositioners between background and data
measurements, the \NA{} of the detection path was reduced with an iris. The
spatial resolution was not impaired by this as we still used the full \NA{} in the
excitation path. After verifying that the only spectral features stem from the
\mose{} exciton at charge neutrality and the attractive polaron at high hole doping, we normalized the 
the low- and high-energy halves of the spectra with respect to the spectra measured at a gate voltage
where there is no signal in the respective energy range. This was done to avoid moving the
nanopositioners between measurements and ensure to have consistent datasets from the same spot.
Due to interference of the multiple reflections at the interfaces in the stack, the lines
in resonant spectra have a dispersive Lorentzian shape. Therefore we use the model
\begin{equation}
f(x) = \Re\left(\frac{a}{\pi} \frac{e^{-i \varphi}}{\sigma - i\left(x - x_0\right)} + c\right)
\end{equation}
with five free parameters $a$, $c$, $\varphi$, $\sigma$, and $x_0$ to fit the spectral
lineshapes. The peak center is at $x_0$, the FWHM is $2\sigma$, the peak height is
$a/(\pi\sigma)$, and the peak area is $a$.

\subsection{Magneto-optical Kerr Effect}

The \moke{} was measured by sending the reflected linearly polarized light through a Wollaston
prism with its optical axis rotated by \SI{45}{\degree} with respect to the polarization
of the excitation beam. For small Kerr rotation angles, this leads to approximately equal
intensities $I_H$ and $I_V$ in the two outgoing beams, which
are measured using two photodiodes. The deviation from \SI{45}{\degree} polarization is
then given by
\begin{equation}
\theta_K = \frac{1}{2}\arcsin\left(
    \frac{I_H - I_V}{I_H + I_V}\right)
    \approx \frac{1}{2}\frac{I_H - I_V}{I_H + I_V}.
\end{equation}

\subsection{Critical Behavior}

To obtain the magnetization and susceptibility curves as a function of temperature, we measure
magnetic field sweeps of the reflection spectra in $\sigma^-$ and $\sigma^+$ polarization at each
temperature and fit the exciton peaks to extract the splitting. Assuming that the splitting is
proportional to the magnetization of \crbr, we get the magnetization curve from the splitting at
zero field and the susceptibility from the slope of the splitting at zero field. We determine the real
zero magnetic field by identifying the center of the hysteresis curve or the point where the two peak
positions cross above the Curie temperature, and find that it is slightly shifted to about
\SI{1}{mT}, probably due to some slightly ferromagnetic component in our setup. For the
magnetization we use the mean splitting of the three points closest to zero field and for the
susceptibility we fit five points around zero field with a linear model to extract the
slope. In both cases, this procedure may systematically underestimate the real value close to
the critical temperature.

\begin{figure}
\includegraphics{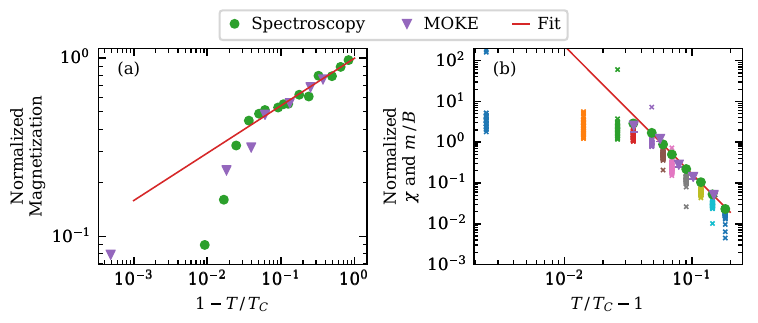}
\caption{(a) Normalized remnant magnetization as function of temperature and (b)
normalized susceptibility as function of temperature. Solid lines, filled circles and
filled triangles are the same as shown in the main text; crosses are $m/B$
from individual data points. In both cases the data points follow the
fitted power law when $\left|T / T_C - 1\right| > 3\cdot10^{-2}$.}
\label{fig:loglog}
\end{figure}

For the simultaneous fits of the magnetization and susceptibility curves, the data
was normalized in order to have similar weights for all data points. Since the susceptibility
was extracted from linear fits to the splitting around zero field, we have a good estimate
for the error on these points that we can include as weighting factors in the fit
for the critical exponents. For the magnetization data points, which correspond directly
to fitted peak positions, the statistical uncertainty is very small but they are subject to
systematic errors, mainly due to thermal expansion that moves the measurement spot on the
sample in the temperature sweep. Both for $T > T_C$ and $T < T_C$ a few
more data points close to $T_C$ were measured that are not shown in Fig.~4 of the main text.
These data points were excluded for three reasons: first, close to $T_C$ it becomes more challenging to extract reliable values due to the vanishing splitting and the diverging susceptibility. Second, we are limited by the minimum step size of the magnetic field. Third, the finite sample size effects and disorder  become dominant close
to the phase transition. In \fig{fig:loglog} we show the full data set
on a logarithmic scale, such that the power laws appear as linear dependence. For
$\left|T/T_C - 1\right| > 3\cdot10^{-2}$ the data points agree well with the fitted power law,
but close to $T_C$ they deviate from it, as pointed out above. Additionally, we show
the measured magnetization divided by the applied magnetic field for all data points above the critical
temperature in \subfig{fig:loglog}{b}, which also follows the same power law.

\bibliographystyle{apsrev4-2}
\bibliography{references}